\documentclass[sigplan,10pt]{acmart}

\usepackage{tikz}
\usepackage{amsmath}

\usepackage{epsfig,endnotes}
\usepackage{graphicx}
\usepackage{booktabs}
\usepackage{longtable}
\usepackage{multicol}
\usepackage{tabularx}
\usepackage{xltabular}
\usepackage{xcolor}
\usepackage{pifont}
\usepackage{float}

\usepackage{listings}
\usepackage{algorithm}
\usepackage{algorithmic}
\usepackage{url}

\usepackage{breakurl}

\usepackage{caption}
\usepackage{enumitem}
\usepackage{endnotes,microtype,xspace,graphicx,fancyvrb,multirow}
\usepackage{thmbox}

\newcommand{\ignore}[1]{}

\newcommand{\codeword}[1]{\textcolor{black}{$\mathsf{#1}$}}

\newcommand{\myparagraph}[1]{\vspace {3pt}\noindent\textbf{\emph{#1}}}

\newcommand{\sys}{IntentOS\xspace}
\newcommand{\arch}{agentic kernel\xspace}
\newcommand{\isyscall}{\codeword{isyscall}\xspace}
\newcommand{\llmbrain}{LLM Planner\xspace}
\newcommand{\agenticplan}{agentic plan\xspace}
\newcommand{\agenticplans}{agentic plans\xspace}
\newcommand{\agentictool}{agentic tool\xspace}
\newcommand{\agentictools}{agentic tools\xspace}
\newcommand{\solutionperformance}{Co-Designed JIT Compilation\xspace}
\newcommand{\solutionsafety}{Safe Micro-Syscall\xspace}
\newcommand{\solutionfault}{Multi-Stage Evolution\xspace}
\newcommand{\topdowndecompose}{Top-Down Decomposition\xspace}
\newcommand{\forkexec}{\codeword{fork+exec}\xspace}
\newcommand{\batchunlink}{\codeword{batch unlink}\xspace}
\newcommand{\slidingwindowmmap}{\codeword{sliding mmap}\xspace}
\newcommand{\llmsidecar}{\codeword{LLM Sidecar}\xspace}
\newcommand{\semanticcomposition}{syscall composition\xspace}
\newcommand{\semanticcompositions}{syscall compositions\xspace}

\DeclareCaptionFormat{mylst}{\hrule#1#2#3}
\usepackage{listings}
\newcommand\YAMLcolonstyle{\color{red}\mdseries}
\newcommand\YAMLkeystyle{\color{black}\small\ttfamily}
\newcommand\YAMLvaluestyle{\color{blue}\mdseries}
\makeatletter
\newcommand\language@yaml{yaml}
\expandafter\expandafter\expandafter\lstdefinelanguage
\expandafter{\language@yaml}
{
  keywords={true,false,null,y,n},
  keywordstyle=\color{darkgray}\bfseries,
  basicstyle=\YAMLkeystyle,                                 
  sensitive=false,
  comment=[l]{\#},
  morecomment=[s]{/*}{*/},
  commentstyle=\color{purple}\ttfamily,
  stringstyle=\YAMLvaluestyle\ttfamily,
  moredelim=[l][\color{orange}]{\&},
  moredelim=[l][\color{magenta}]{*},
  moredelim=**[il][\YAMLcolonstyle{:}\YAMLvaluestyle]{:},   
  morestring=[b]',
  morestring=[b]",
  literate =    {---}{{\ProcessThreeDashes}}3
                {>}{{\textcolor{red}\textgreater}}1     
                {|}{{\textcolor{red}\textbar}}1 
                {\ -\ }{{\mdseries\ -\ }}3,
}
\lst@AddToHook{EveryLine}{\ifx\lst@language\language@yaml\YAMLkeystyle\fi}
\makeatother
\newcommand\ProcessThreeDashes{\llap{\color{cyan}\mdseries-{-}-}}

\begin{document}


\title{
An Agentic OS Kernel for Flexible and Enhanced Semantics
}

\author{Haocheng Wang}
\affiliation{
  \institution{Shanghai Jiao Tong University}
  \city{Shanghai}
  \country{China}
}
\email{hcwang369@sjtu.edu.cn}

\author{Xinyue Sun}
\affiliation{
  \institution{Shanghai Jiao Tong University}
  \city{Shanghai}
  \country{China}
}
\email{sxxyyy2005@sjtu.edu.cn}

\author{Hao Yi}
\affiliation{
  \institution{Zhejiang University}
  \city{Hangzhou}
  \country{China}
}
\email{3240104995@zju.edu.cn}

\author{Dong Du\,\texorpdfstring{\textsuperscript{\ding{41}}}{}}
\affiliation{
  \institution{Shanghai Jiao Tong University}
  \department{Advanced Research Center for Agent-Oriented OS}
  \city{Shanghai}
  \country{China}
}
\email{dd_nirvana@sjtu.edu.cn}

\author{Yubin Xia}
\affiliation{
  \institution{Shanghai Jiao Tong University}
  \department{Advanced Research Center for Agent-Oriented OS}
  \city{Shanghai}
  \country{China}
}
\email{xiayubin@sjtu.edu.cn}

\author{Haibo Chen}
\affiliation{
  \institution{Shanghai Jiao Tong University}
  \city{Shanghai}
  \country{China}
}
\email{haibochen@sjtu.edu.cn}

\renewcommand{\shortauthors}{Wang et al.}

\begin{abstract}

Modern OS kernels expose general-purpose but rigid interfaces (e.g., syscalls) that are slow to evolve due to compatibility, performance, and security constraints,
leaving apps to pay for mismatched semantics and legacy APIs.
Existing kernel mechanism customization approaches (e.g., eBPF) improve flexibility but still face strict safety and performance restrictions and thus require substantial expertise. What's more, they targets customized kernel mechanisms rather than kernel abstractions.

This paper presents \arch, a new architecture that decomposes monolithic syscall functions into fine-grained, composable kernel-native \emph{micro-syscalls}, and then enables Large Language Models (LLMs) to orchestrate them dynamically.
Agentic kernel introduces a new entry, \isyscall (intent syscall), which lets apps submit high-level \emph{intents} describing their requirements.
Then the kernel leverages LLMs to resolve each intent into composition plan of micro-syscalls to satisfy the user's goal. 
LLMs can further optimize the plan continuously at kernel runtime.
To eliminate LLM inference costs on the critical path, \arch caches and reuses verified plans to bypass LLM entirely.
Crucially, \arch's safety guarantees are independent of LLM correctness: micro-syscalls follows syscall-level security mechanisms, ensuring OS safety under arbitrary combinations and orchestration.

We implement \arch in Linux, called \sys, and evaluate it across four kernel versions and three application domains (about ten intents each).
\sys outperforms baseline Linux in real-world applications, achieving up to 17.55\% latency reduction for backend processes creation, 6.97\% performance gain for multiple files operations, and 75.52\% lower performance impact for frontend Nginx server during memory operations triggered by backend tasks.

\end{abstract}	

\settopmatter{printacmref=false}
\setcopyright{none}
\renewcommand\footnotetextcopyrightpermission[1]{}
\settopmatter{printfolios=true}
\maketitle
\pagestyle{plain}

\section{Introduction}
\label{sec:intro}

\begin{figure}[t]
  \centering
  \begin{minipage}[t]{0.48\linewidth}
    \centering
    \includegraphics[width=1.07\linewidth]{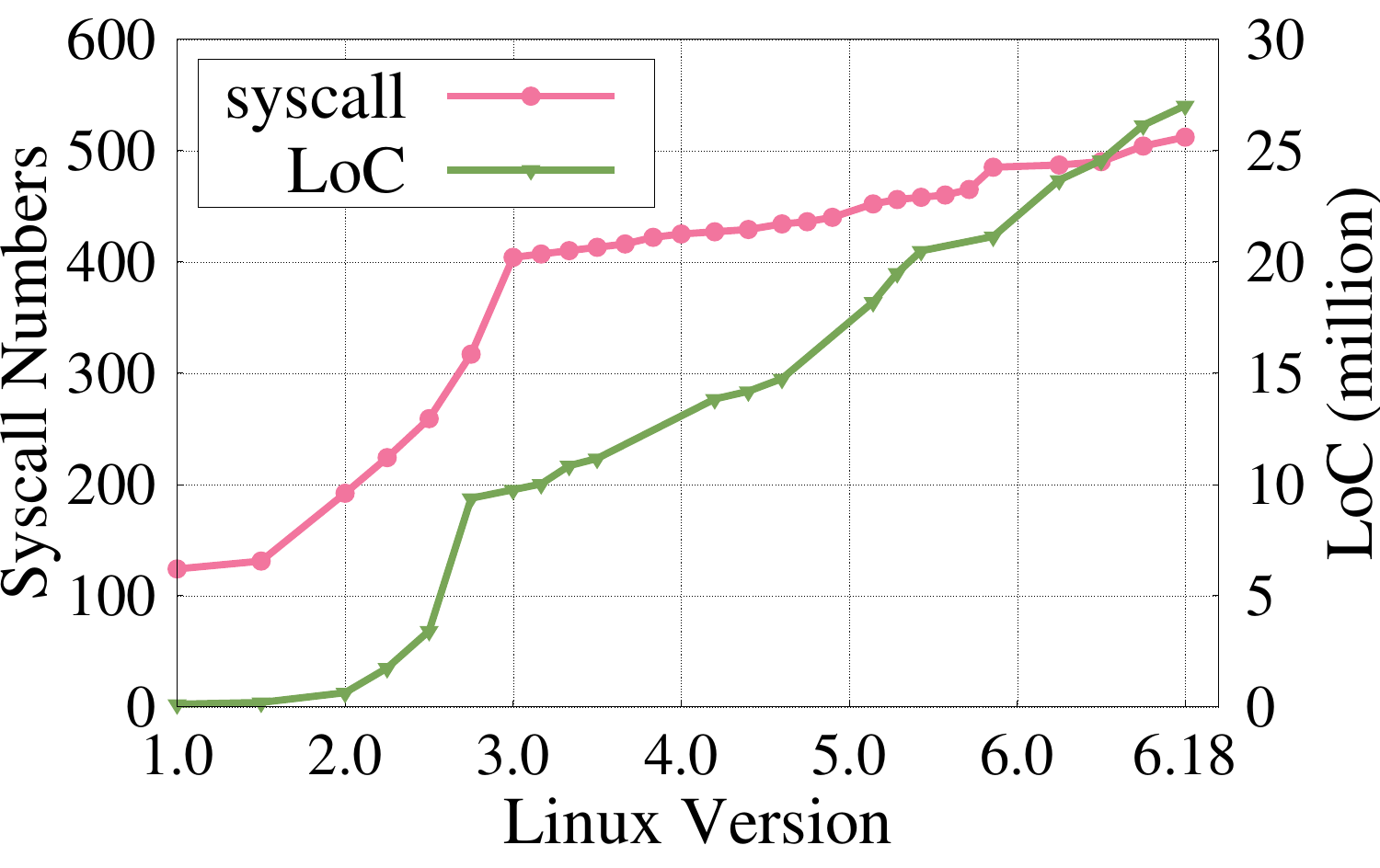}
    \footnotesize
    \textbf{(a) Syscall Count Evolution.}
  \end{minipage}
  \hfill
  \begin{minipage}[t]{0.48\linewidth}
    \centering
    \includegraphics[width=1.07\linewidth]{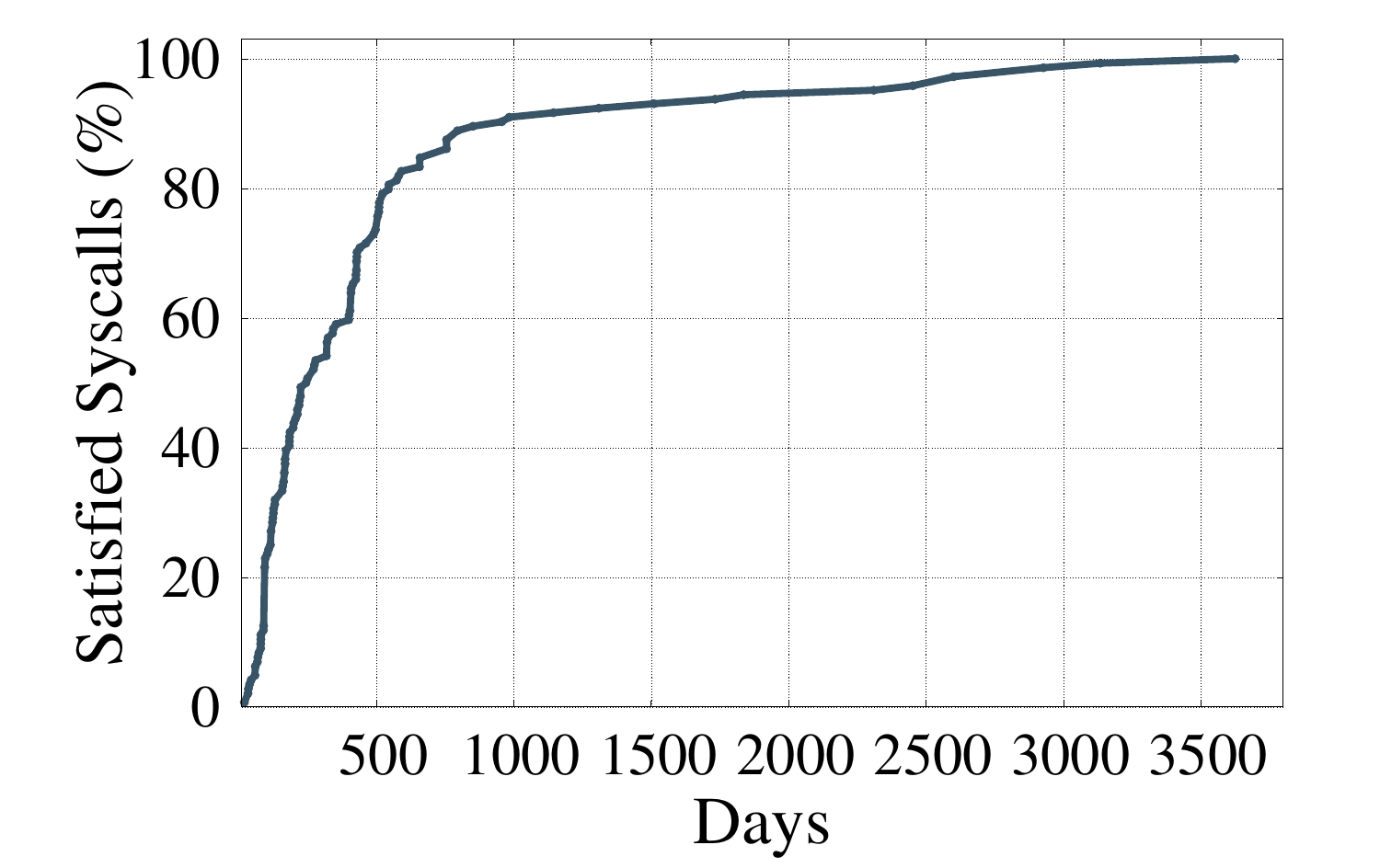}
    \footnotesize
    \textbf{(b) Proposal-to-Merge Time CDF.}
  \end{minipage}
  \setlength{\abovecaptionskip}{3pt}
  \caption{\textbf{Syscall evolution in Linux.} (a) shows growth of syscall and LoC across Linux versions, and (b) shows the CDF of proposal-to-merge time for new syscalls since v2.6.}
  \label{fig:syscall-evolution}
\end{figure}

Operating systems (OSes) mediate access to hardware resources by exposing a set of semantics (i.e., syscalls~\cite{syscall2-manual}) for user-mode programs to request system services.
Due to generality and compatibility concerns~\cite{10.1145/2901318.2901350}, existing syscalls are notoriously fixed and slow to evolve.
As shown in \autoref{fig:syscall-evolution}, from Linux 3.0 to Linux 6.18, although the lines of code (LoC) nearly tripled, the number of syscalls increased by only 26.7\%.
Furthermore, analysis of the time from proposal (on the mailing list, since v2.6) to mainline merge reveals that nearly 40\% of new syscalls take over 400 days to be adopted.
They highlight the significant engineering effort required to introduce new abstractions into the kernel.
Consequently, applications often compose multiple syscalls to achieve complex intents, which we define as \emph{\semanticcomposition}.

Applications often endure sub-optimal performance due to rigid kenrel interfaces and fragmented semantics in \semanticcomposition.
For example, the classic \codeword{fork} and \codeword{execve} pattern in Linux is routinely composed to spawn a new process.
Beyond other well-known issues~\cite{10.1145/3317550.3321435}, this pattern performs redundant work:
\codeword{fork} unconditionally copy \codeword{mm\_struct} for child, 
only for the subsequent \codeword{execve} to promptly discard it 
and allocate a new one via \codeword{mm\_alloc} sub-function.
This inefficiency persists because both syscalls operate independently with partial semantics, oblivious to the application's overarching intent.
Although alternatives like \codeword{vfork}~\cite{syscall-vfork} avoid duplicated copy, they impose restrictions (e.g., blocking the parent) that
brings new safe and performance issue.

To mitigate sub-optimal \semanticcomposition, researchers have explored a wide spectrum of customization mechanisms and OS architectures.
EBPF~\cite{ebpf, 308758, 280870, 10.1145/3731569.3764820, 298730, 10.1145/3731569.3764796, 10.1145/3694715.3695950, 10.1145/3689031.3717497, 10.1145/3689031.3696094, 10.1145/3627703.3629562, 308448, 308452, 298490, 298531, 10.1145/3620666.3651387} enables safe in-kernel extensions at predefined hooks. But it is not designed to introduce customized system calls; 
Other approaches, such as exokernels~\cite{engler1995exokernel}, LibOS-related works~\cite{180274, 10.1145/3689031.3717490, 10.1145/3503222.3507759}, unikernels~\cite{10.1145/2499368.2451167}, Skyloft~\cite{10.1145/3694715.3695973}, and DemiKernel~\cite{10.1145/3477132.3483569}, shift functionality out of the kernel to give apps more control over system services.
However, both maintaining and using the above frameworks require expert knowledge, limiting their practical usability. 
What's more, they focus on customized behaviors defined during development-time, missing the capability to discover and optimize hot \semanticcomposition at runtime.

The key gap is that, due to the diversity of apps' intents, syscalls' fixed semantics limit the flexibility and performance of \semanticcomposition, while nowadays customization mechanisms (e.g., eBPF) mainly target \emph{customized kernel mechanisms} rather than \emph{kernel abstractions and interfaces}.

This paper presents \arch, a novel kernel architecture that enables applications to build flexible and enhanced kernel semantics including runtime capabilities, without recompiling the kernel.
To provide more flexible optimization opportunities, \arch derives fine-grained sub-functions, termed \emph{micro-syscalls}, from coarse-grained syscall functions.
Applications can compose appropriate micro-syscalls within a single kernel trap, thereby reducing redundant operations in syscall compositions; 
To shield applications from the complexity of micro-syscall orchestration and enable dynamic decision at runtime, \arch introduces an LLM-powered interaction paradigm, \emph{intent as a syscall}. It provides a new interface, \isyscall (intent call), to express user's intent for \semanticcomposition with flexible hint. And recent advances in LLMs make automated plan generation feasible: given \emph{micro-syscalls}, LLMs can compose them for classic POSIX semantics and even discover new performance optimizations. And runtime metrics collected in the kernel can help LLMs to build more appropriate implementation. 

This motivates us to leverage LLMs as the planning backend (\llmbrain) to generate execution plans (\agenticplans) from applications' intent. However, it is non-trivial to realize \arch.

\myparagraph{Challenge-I: Latency of runtime plan generation.}
Runtime plan generation introduces latency on the critical path of applications:
 prompt-based intents must be delivered to LLM, and applications have to wait for the result.
LLM inference always consumes tens of seconds, even minutes for complex intents~\cite{oracle-llama3-70b-benchmark}, while syscall latency is usually on the order of microseconds to milliseconds. 
The reasoning overhead can easily overshadow the performance gains brought by optimizing \semanticcompositions, and may even cause severe negative optimization.

\myparagraph{Challenge-II: Correctness of automatically generated plans.}
Automatically generated plans may be incorrect when planner faces complex and ambiguous hints.
Arbitrary decomposition of syscalls without unified methodology leads to the dilemma:
Fine-grained \agentictools derived from syscall functions will lose the original context and have unclear functional boundaries (e.g., functional deficiency).
LLMs may hallucinate under ambiguous hints~\cite{healy2026internalrepresentationsindicatorshallucinations,kim2025detectingllmhallucinationlayerwise,wiggers2025shortanswershallucinations, 10.1145/3703155}, frequently generating inaccurate \agenticplans that undermine kernel robustness and reliability.

\myparagraph{Challenge-III: Kernel safety under arbitrary plan compositions.}
Without safety guarantee, the fine-grained \agentictools can be combined arbitrarily. 
Regardless of whether a plan is generated by LLMs or crafted by a developer, malicious users~\cite{maloyan2026promptinjectionattacksagentic} have the chances to exploit fine-grained \agentictools to cause kernel panics (e.g., error arguments for undefined behavior), perform privileged operations (e.g., killing non-child processes), or access and corrupt kernel data that should not be exposed to them.

This paper presents \sys, an OS kernel that realizes \arch. 
To address the three challenges, \sys introduces three key techniques.

First,
\sys introduces \textbf{\solutionperformance} to cache and reuse \agenticplans.
For unseen intents, kernel will pass them to the \llmbrain to generate \agenticplans. \llmbrain will return well-formatted plan that can be executed to finish requirements.
For hot intents, kernel will further cache their plans, so the app can bypass LLM inference completely when encountering them again. 

Second,
\sys adopts \textbf{\solutionfault} mechanism to narrow the decision space and improve inference accuracy gradually.
We refer to top-down philosophy to decompose the syscall functions into fine-grained, multi-level and functionally coherent micro-syscalls.
Parent functions and their derived sub-functions together form a multi-level hierarchy.
\sys will strictly construct initial \agenticplans following syscall-grained tools at highest-level to guarantee basic correctness and accuracy.
And then substitute it with low-level sub-tools for optimization.

Last,
\sys proposes \textbf{\solutionsafety} to ensure the security of \agentictools.
Similar to syscalls, each tool exists as a self-contained function that can be directly invoked by users. 
\sys performs pre-execution security checking to avoid malicious parameter values and unauthorized operations. 
All kernel resources associated with a micro-syscall remain protected within kernel space, and are exposed to users only via object handles.
Lock reentrancy guarantees and recovery mechanisms further prevent the kernel from becoming stuck in corrupted states.

\begin{figure*}[tb]
    \setlength{\belowcaptionskip}{-10pt}
  \setlength{\abovecaptionskip}{-10pt}
  \centering
  \includegraphics[width=0.99\linewidth]{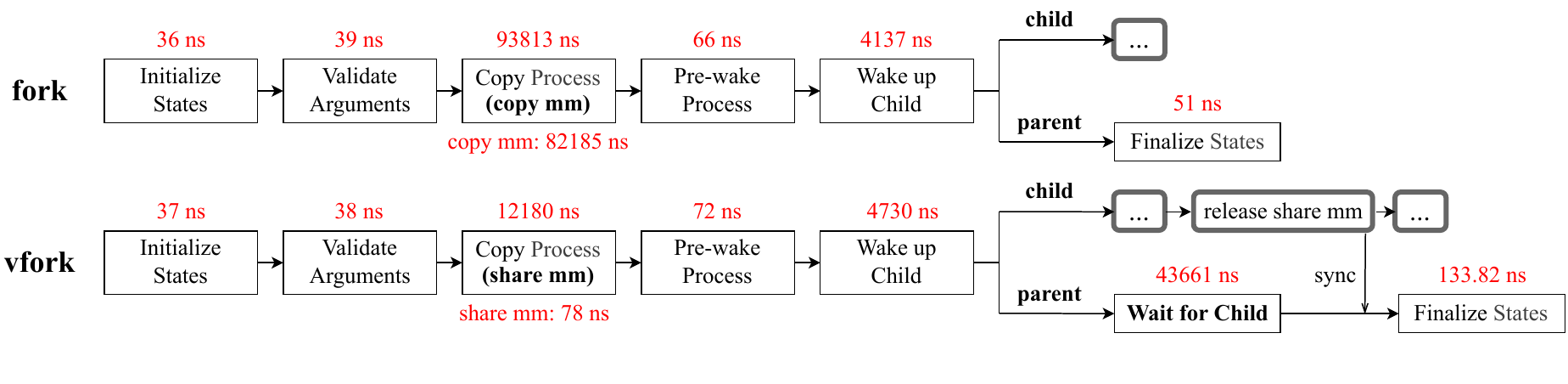}
  \caption{\textbf{Fork/vfork operations breakdown.}}
  \label{fig:fork-vfork-breakdown}
\end{figure*}

\begin{table}[htbp]
  \centering
  \setlength{\belowcaptionskip}{-5pt}
  \setlength{\abovecaptionskip}{0pt}
  \scriptsize
  \caption{\textbf{Shared sub-functions across process management syscalls in Linux.} The rigid bundling of these sub-functions into fixed syscalls leads to redundant operations when syscalls are used in sequence (e.g., \codeword{fork} + \codeword{execve}).}
  \begin{tabular}{lccccc}
    \toprule
    \textbf{Kernel Sub-function} & \textbf{\texttt{fork}} & \textbf{\texttt{vfork}} & \textbf{\texttt{clone}} & \textbf{\texttt{execve}} & \textbf{\texttt{unshare}} \\
    \midrule
    \texttt{dup\_task\_struct()} & Yes & Yes & Yes & No & No \\
    \texttt{copy\_files()} / \texttt{dup\_fd()} & Yes & Yes & Yes & Yes* & Yes* \\
    \texttt{copy\_fs()} & Yes & Yes & Yes & No & Yes \\
    \texttt{copy\_sighand()} & Yes & Yes & Yes & Yes\dag & No \\
    \texttt{copy\_mm()} / \texttt{dup\_mm()} & Yes & Share & Cond. & No & No \\
    \texttt{mm\_alloc()} & No & No & No & Yes & No \\
    \bottomrule
    \multicolumn{6}{l}{\scriptsize * via \texttt{unshare\_files()} $\rightarrow$ \texttt{dup\_fd()}; \dag via \texttt{unshare\_sighand()}}
  \end{tabular}
  \label{tab:syscall-sub-function}
\end{table}

We implement \sys on Linux, and evaluate its basic inference accuracy across three system domains (file system, memory management, and communication), each with ten representative intents.
We conduct compatibility tests across four popular kernel versions (Linux 3.18, 4.19, 5.15 and 6.18) to show that high-level intents remain compatible with the evolution of kernel.
We evaluate the performance of optimized \agenticplans using three cases: \forkexec, \batchunlink, and \slidingwindowmmap, each with both microbenchmarks and end-to-end application tests. 
In the \forkexec test with busybox-ash, \sys achieves up to 17.55\% latency reduction; 
In the \batchunlink test with Unix rm, \sys achieves up to 6.97\% performance gain; 
In the \slidingwindowmmap test with backend file-grep, \sys reduces the performance impact to the frontend Nginx service by up to 75.52\%.
\sys will be fully open-sourced.

\section{How OS Kernel Abstractions Evolve?}

\label{s:motiv}

Application performance significantly depends on the exposed OS abstractions and interfaces. 
Different kernel semantic design choices lead to vastly different system properties and application performance. 
However, there is no silver bullet; unsuitable kernel semantics can lead to error function, severe performance degradation~\cite{zhang2019m, 10.1145/2699681, 10.1145/3132747.3132780}, security vulnerabilities~\cite{10.1145/3458336.3465273, 10.1145/3445814.3446728}, and high engineering costs. 
As hardware becomes more heterogeneous~\cite{baumann2009multikernel} and applications more diverse (many vibe coding apps), the ``one-size-fits-all'' abstraction provided by traditional kernels is increasingly becoming a bottleneck~\cite{engler1995exokernel, 10.1145/3477132.3483569}.
The evolution of kernel semantics is important to satisfy the specific requirements of various applications.

\subsection{Kernel Semantic Evolution Dilemma}
 
For reasons of compatibility and subsequent maintainability~\cite{lwn-bpf-mm}, adding new kernel semantics takes a long time, and the overall trend is slowing down.
We present a thorough analysis of how kernel semantics evolve over time:
First, the number of new syscalls increases slowly. As shown in \autoref{fig:syscall-evolution}(a), over a 14-year span from Linux 3.0 to 6.18, the number of system calls increases by only 26.7\%, while the lines of code nearly triple. 
Secondly, the new syscall always takes a long time to merge into the Linux mainline. As shown in \autoref{fig:syscall-evolution}(b), nearly 40\% of new syscalls take more than 400 days to merge. Some syscalls like \codeword{clone3} even take nearly a decade from the initial proposal in the mailing list to being released in a kernel version~\cite{clone3-rfc, syscall2-manual}.

\myparagraph{\underline{Implication-1:}}
\textbf{The slow iteration imposes a high burden on evolving kernel semantics and suggests that, while OS mechanisms evolve rapidly, abstraction evolution is much slower.}
Syscalls are designed for general-purpose usage, but not for specific intents. 
As a result, applications often compose one or more syscalls 
to implement complex functionalities, which we call \emph{\semanticcompositions}.

\begin{table*}[tb]
\begingroup
\centering
\setlength{\belowcaptionskip}{0pt}
\setlength{\abovecaptionskip}{-8pt}
\fontsize{6.6}{7.2}\selectfont
\setlength{\tabcolsep}{0pt}
\renewcommand{\arraystretch}{1.06}
\setlength{\extrarowheight}{0.25pt}
\def\syscombfuncwidth{0.48}
\def\syscombpatternwidth{0.49}
\def\syscombtablelabel#1{{\fontsize{8.6}{9.2}\selectfont\textbf{#1}}\par\vspace{3pt}}
\def\syscombhead#1{\textbf{#1}}
\def\syscombrow#1#2{%
  \textbf{\texttt{#1}} & #2 \\
}
\def\syscombrowrule{\specialrule{0.15pt}{0.45pt}{0.45pt}}
\begin{minipage}[t]{0.326\linewidth}
  \centering
  \syscombtablelabel{(a) GCC}
  \begin{tabular}{@{}>{\raggedright\arraybackslash}p{\syscombfuncwidth\linewidth}@{\hspace{3pt}}>{\raggedright\arraybackslash}p{\syscombpatternwidth\linewidth}@{}}
    \toprule
    \syscombhead{Func} & \syscombhead{Combination Pattern} \\ \midrule
    \syscombrow{pex\_input\_file}{\texttt{open} + \texttt{close} + \texttt{open}}
    \syscombrowrule
    \syscombrow{close\_at\_file}{\texttt{open} + \texttt{write} + \texttt{close}}
    \syscombrowrule
    \syscombrow{execute}{\texttt{pipe} + \texttt{vfork} + \texttt{dup2} + \texttt{execve} + \texttt{waitpid}}
    \syscombrowrule
    \syscombrow{pex\_unix\_exec\_child}{\texttt{pipe2} + \texttt{vfork} + \texttt{dup2} + \texttt{close} + \texttt{execvp}}
    \bottomrule
  \end{tabular}
\end{minipage}
\hfill
\begin{minipage}[t]{0.326\linewidth}
  \centering
  \syscombtablelabel{(b) Nginx}
  \begin{tabular}{@{}>{\raggedright\arraybackslash}p{\syscombfuncwidth\linewidth}@{\hspace{3pt}}>{\raggedright\arraybackslash}p{\syscombpatternwidth\linewidth}@{}}
    \toprule
    \syscombhead{Func} & \syscombhead{Combination Pattern} \\ \midrule
    \syscombrow{ngx\_file\_info\_wrapper}{\texttt{open} + \texttt{fstat} + \texttt{close}}
    \syscombrowrule
    \syscombrow{ngx\_open\_and\_stat\_file}{\texttt{open} + \texttt{fstat} + \texttt{fcntl}}
    \syscombrowrule
    \syscombrow{ngx\_linux\_\allowbreak sendfile\_chain}{\texttt{setsockopt} + \texttt{sendfile}}
    \syscombrowrule
    \syscombrow{ngx\_udp\_connect}{\texttt{socket} + \texttt{ioctl} + \texttt{connect} + \texttt{epoll\_ctl}}
    \bottomrule
  \end{tabular}
\end{minipage}
\hfill
\begin{minipage}[t]{0.326\linewidth}
  \centering
  \syscombtablelabel{(c) SQLite}
  \begin{tabular}{@{}>{\raggedright\arraybackslash}p{\syscombfuncwidth\linewidth}@{\hspace{3pt}}>{\raggedright\arraybackslash}p{\syscombpatternwidth\linewidth}@{}}
    \toprule
    \syscombhead{Func} & \syscombhead{Combination Pattern} \\ \midrule
    \syscombrow{unixShmMap}{\texttt{fstat} + \texttt{pwrite} + \texttt{mmap}}
    \syscombrowrule
    \syscombrow{fcntlSizeHint}{\texttt{fstat} + \texttt{fallocate} + \texttt{mmap}}
    \syscombrowrule
    \syscombrow{unixOpenSharedMemory}{\texttt{fstat} + \texttt{open} + \texttt{fchown} + \texttt{fcntl} + \texttt{ftruncate}}
    \syscombrowrule
    \syscombrow{unixSync}{\texttt{fdatasync} + \texttt{open} + \texttt{fdatasync} + \texttt{close}}
    \bottomrule
  \end{tabular}
\end{minipage}

\par\vspace{8pt}
\caption{\textbf{Syscall combinations on different applications.} Each sub-table summarizes representative syscall compositions for a different application. The Func column lists concrete function names that appear in the corresponding source code.}
\label{tab:syscall-combination-on-diff-app}
\endgroup
\end{table*}

\begin{figure}[t]
  \centering
  \includegraphics[width=0.75\linewidth]{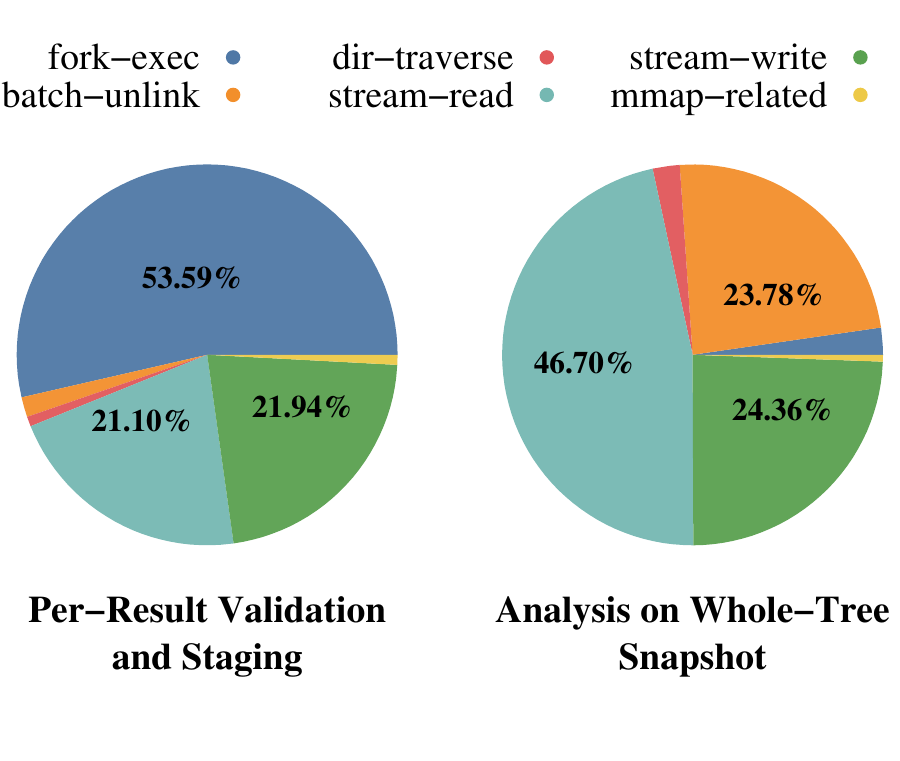}
  \setlength{\belowcaptionskip}{-10pt}
  \setlength{\abovecaptionskip}{-5pt}
  \caption{\textbf{Distribution of \semanticcompositions under different runtimes.} Comparison of six \semanticcompositions in two runtime workloads. Tests are conducted with Bash and Unix coreutils-related applications.}
  \label{fig:syscall-combination-diff-runtime}
\end{figure}

\subsection{The Sub-optimality of Rigid Syscall}
\label{s:motiv-case-analysis}

The evolution dilemma makes system semantics rigid, leading to performance degradation in \semanticcompositions.
To deeply understand it, we analyze the process creation mechanism in Linux
with \codeword{fork} and \codeword{execve} related syscalls.

In Linux, syscalls can be decomposed into a set of sub-functions. 
E.g., the \codeword{fork} syscall relies \codeword{copy\_process} function, which sequentially invokes sub-function like \codeword{copy\_mm}, \codeword{copy\_files}, and \codeword{copy \_sighand}. 
As shown in \autoref{tab:syscall-sub-function}, different syscalls reuse these sub-functions to provide specific semantics with fixed combinations, which carry unnecessary overhead at specific scenario.

When spawning a subprocess, the classic approach requires a sequential combination of \codeword{fork} and \codeword{execve}, which includes redundant operations:
In \codeword{fork}, the kernel blindly invokes \codeword{copy\_mm} and \codeword{dup\_mm} to create a full Copy-on-Write (CoW) duplicate of the parent's address space. 
This involves allocating a new \codeword{mm\_struct} and copying all PTEs, which consumes many cycles~\cite{10.1145/3447786.3456258}. 
However, the newly-copied objects will be discarded instantly by \codeword{exec\_mmap} when child calls \codeword{execve}.
Although this redundant memory copying can be eliminated in \codeword{vfork} by sharing \codeword{mm\_struct}, it introduces other redundant operations: 
in order to avoid memory data corruption caused by concurrent access, parent has to be pended in \codeword{vfork} until child releases shared \codeword{mm\_struct}.
These redundant operations waste significant CPU cycles and degrade performance. As illustrated in \autoref{fig:fork-vfork-breakdown}, the \codeword{copy\_mm} operation in \codeword{fork} accounts for 78.89\% of the latency, and the waiting operation in \codeword{vfork} accounts for 70.281\% of the latency. 
In \S\ref{sec:case-study}, we show that the redundant operations can be significantly reduced by flexibly composing underlying sub-functions.

\myparagraph{\underline{Implication-2:}}
\textbf{Syscall compositions are sub-optimal for specific scenarios due to their fixed internal implementations and the inability to flexibly compose underlying kernel primitives}. 

\subsection{Development-Time Customizations Fall Short}
Researchers have proposed many approaches at development-time: 
Mechanisms like eBPF allow applications to inject custom logic into the kernel safely, enabling efficient kernel abstractions for networking and observability~\cite{cilium}. And \codeword{io\_uring} provides a high-performance asynchronous I/O interface with flexible capabilities~\cite{iouring}.
Although they provide the chance to customize kernel at development-time, 
they are not enough for optimizing \semanticcompositions.

We present a study on how apps will use \semanticcompositions, taking gcc, nginx, and sqlite as examples.
As shown in \autoref{tab:syscall-combination-on-diff-app}, each app has own requirements and thus selects different \semanticcompositions. 
If the kernel provide a dedicated optimized interface for each composition, the number of them would quickly explode. 
The problem is further complicated by the fact that the effectiveness of many optimizations depends on runtime states. 
As shown in \autoref{fig:syscall-combination-diff-runtime}, hot \semanticcompositions may vary across workloads, applications has no runtime information for optimization at development-time. 
Moreover, some optimizations are beneficial only under specific runtime states. 

\myparagraph{\underline{Implication-3:}}
\textbf{Different applications induce different \semanticcompositions, which cannot be exhaustively enumerated during development.}
These limitations indicate that kernels need the ability to discover and optimize \semanticcompositions \textbf{at runtime}. Unfortunately, current kernels still lack this ability for two reasons: 
First, syscall invocations are exposed to the kernel as discrete events, making it difficult for the kernel to identify hot \semanticcompositions or to optimize them as a new whole semantic.
Second, syscall functions is too coarse-grained to be flexibly reorganized at runtime. Kernel cannot decompose them into smaller logic units and re-orchestrate them, which misses many optimization opportunities.

\begin{figure}[tb]
  \centering
  \includegraphics[width=0.99\linewidth]{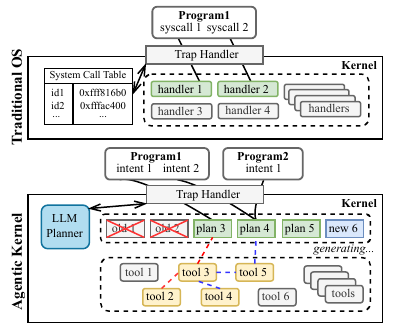}
  \setlength{\belowcaptionskip}{-10pt}
  \setlength{\abovecaptionskip}{-10pt}
  \caption{\textbf{OS architecture comparison.} \sys converts the interface from API to intent, thus enabling the ability to compose tools flexibly.}
  \label{fig:arch-compare}
\end{figure}

\subsection{Opportunities: Composable Kernel Primitives}
The analysis above reveals that if syscall internals can be decomposed as composable primitives, many optimizations would be possible.
This suggests a different kernel design philosophy: except for exposing monolithic syscalls, the kernel also expose sub-functions that can be composed in one kernel trap, which is called micro-syscall. 
The remaining question is: who composes these sub-functions?
Manual composition shifts the burden back to developers and still lack the global runtime information of the kernel.
Recent advances in LLMs with tool-use capabilities~\cite{xi2023risepotentiallargelanguage} offer an answer: given tool descriptions, LLMs can understand high-level intents and reason to generate correct compositions.
For example, in software engineering, AI agents~\cite{yang2024sweagentagentcomputerinterfacesenable} can autonomously implement features by navigating codebases and selecting appropriate actions.
We validate this claim by showing that LLMs, given our micro-syscall decomposition, can independently re-derive classic POSIX semantics and optimized variants (\S\ref{sec:case-study}).

\begin{figure}
  \centering
  \includegraphics[width=1.0\linewidth]{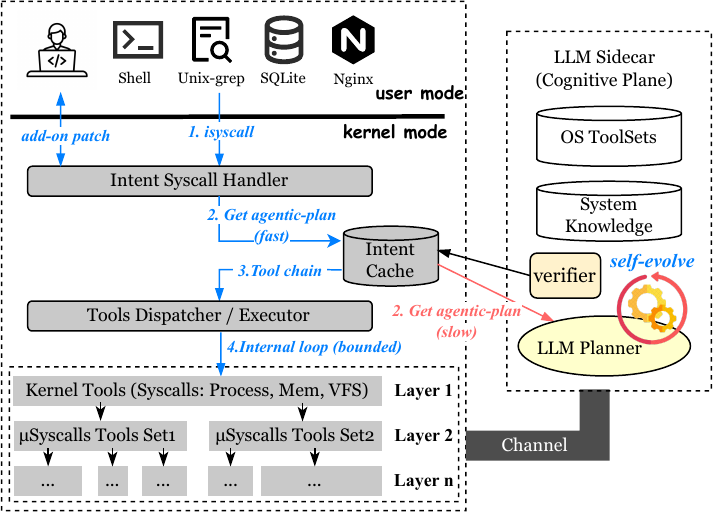}
  \setlength{\belowcaptionskip}{-15pt}
  \setlength{\abovecaptionskip}{-5pt}
  \caption{\textbf{\sys architecture.}}
  \label{fig:design-arch}
\end{figure}

\subsection{Agentic Kernel: Micro-Syscall Decomposition and LLM Runtime Orchestration}

We propose \textbf{\arch}, a novel OS architecture that enables runtime composition of kernel services through fine-grained, composable kernel semantics.
We highlight two progressive shifts from traditional kernels to \arch, as shown in \autoref{fig:arch-compare}.
First, to provide infrastructure for LLM orchestration, we decompose coarse-grained syscall function into fine-grained sub-functions, namely micro-syscalls, to expose more internal optimization opportunities of \semanticcompositions. 
Second, to liberate developers from complex micro-syscalls orchestration and take kernel runtime state into account, we impose LLMs as planner backend. 
Instead of invoking discrete semantics that dictate how a task should be done (e.g., \codeword{socket}, \codeword{bind}, \codeword{listen}), apps issue an \isyscall (intent syscall) that specifies what they want to achieve (e.g., \textit{``open files a, b, and c...''}).
Kernel resolves intents into execution plans dynamically at runtime with the assistant of LLMs.
Natural language intent provides flexible and enhanced semantics, enabling the kernel to efficiently adopt appropriate optimizations.

We build an OS prototype of \arch based on Linux, called \sys, which embodies the core ideas of \arch as well as the unique kernel characteristics of Linux.
Notably, \arch is designed to enhance, not replace, existing ecosystems. It aims to provide a seamless evolution path for modern computing systems.

\section{Overview of \sys}
\label{sec:overview}

The architecture of \sys is shown in \autoref{fig:design-arch}.

\myparagraph{Intent Syscall Entry Point.}
\sys provides the \textit{intent syscall} (i.e., \isyscall) as entry point, shifting the kernel interface from imperative commands to declarative intents.
As illustrated in \autoref{fig:isyscall-comparison}, traditional POSIX paradigm requires the application to manually orchestrate \semanticcompositions and handle the lifecycle split between parent and child processes.
In contrast, \isyscall allows the application to submit natural language strings describing the requirements.
Kernel then takes full responsibility for communicating with \llmbrain, generating safe execution plans, and managing the micro-syscall sequencing and session branching internally.
Moreover, \isyscall decouples actual args from the intent prompt so that the \agenticplans can be reused with changed args.

\begin{figure}[t]
  \centering
  \setlength{\belowcaptionskip}{0pt}
  \setlength{\abovecaptionskip}{-2pt}
  \lstdefinestyle{densecode}{
    language=C,
    basicstyle=\ttfamily\scriptsize,
    keywordstyle=\color{blue},
    commentstyle=\color{green!50!black},
    stringstyle=\color{teal},
    numbers=none,
    frame=none,
    xleftmargin=0pt,
    xrightmargin=0pt,
    aboveskip=0pt,
    belowskip=0pt,
    lineskip=-1pt,
    showstringspaces=false,
    tabsize=2,
    breaklines=true,
    columns=flexible,
    keepspaces=true,
  }
  \begin{minipage}[t]{0.49\linewidth}
    \vspace{0pt}
    {\color{red}\scriptsize\textbf{$\blacktriangleright$ Traditional POSIX}}\\[-11pt]
    \begin{lstlisting}[style=densecode]
void backup_config() {
  pid_t pid = fork();
  if (pid < 0) exit(1);
  if (pid == 0) {
    close(1);
    open("config.bak", O_CREATE|O_WRONLY);
    char *argv[] = {
      "cat", "config", 0};
    exec("cat", argv);
  } else {
    waitpid(pid);
    printf("Backup complete.\n");
  }
}
    \end{lstlisting}
  \end{minipage}%
  \hfill
  \begin{minipage}[t]{0.49\linewidth}
    \vspace{0pt}
    {\color{red}\scriptsize\textbf{$\blacktriangleright$ Agentic Kernel}}\\[-11pt]
    \begin{lstlisting}[style=densecode]
void backup_config() {
  const char *intent =
    "Create a child process, then follow split paths: "
    "In child: copy file from arg1 to arg2 and finish. "
    "In parent: set the child pid in user space ptr arg0.";
  pid_t pid;
  uint64_t args_list[3];
  args_list[0] = &pid;
  args_list[1] = "config";
  args_list[2] = "config.bak";
  isyscall(intent, args_list);
  waitpid(pid);
  printf("Backup complete.\n");
}
    \end{lstlisting}
  \end{minipage}
  \caption{\textbf{Comparative implementation of a backup task in POSIX and \sys.} \sys handles the semantic complexity of the flexible intent internally through its \llmbrain-driven tools orchestrator, reducing applications' effort to convert the intent to system calls manually.}
  \label{fig:isyscall-comparison}
\end{figure}

\myparagraph{\llmbrain in Sidecar.}
In \sys, \llmbrain is encapsulated in an independent component called \llmsidecar. \llmsidecar bridges the gap between the kernel execution environment and LLM-oriented context.
It maintains sessions for every processes, while also managing the toolset registered by the kernel. Requests to build a plan from intent will be sent to LLMs through it. 
\llmsidecar communicates with the kernel through dedicated privileged channel, ordinary users can only interact with \llmbrain and enable new semantics only through \isyscall-related entries. 
\llmsidecar serves only as a external decision maker without execution capability. Actual execution is always performed on the kernel side. 

\myparagraph{Hierarchical Kernel-native Tool Set.}
\sys organizes all orchestratable units(syscalls, micro-syscalls) as \agentictools in multi-level structure. 
It wraps syscalls functions to support basic kernel semantics (Layer-1 tools).
Besides, based on syscall-based tools, \sys provides a set of fine-grained micro-syscall \agentictools to support optimized \agenticplans. 
These \agentictools are organized in a top-down hierarchy, where a tool at upper level can be decomposed into multiple sub-tools.
Each \agentictool consists of two parts: In \llmsidecar, it registers the tool schema in the LLM inference context; 
In OS kernel, it implements the actual function of the tool with micro-syscall safety mechanism (\S\ref{sec:micro-syscall}).

\myparagraph{Robust Execution Plane with Plan Cache.} 
Execution plane (left side of \autoref{fig:design-arch}) forms the major body of \sys.
It is responsible for efficiently handling declarative intents from applications, managing the lifecycle of execution plans, and robustly orchestrating tool following inferred plans. 

Execution plane introduces a dual-path intent handling mechanism: a fast path for intents with cached plans and a slow path for new intents.
What makes dual-path execution practical is the args-decoupled \agenticplan,
together with its just-in-time (JIT) construction and cache (\S\ref{sec:cache}).
For unseen intents, execution plane has to use the slow path, forwarding the request to the \llmsidecar.
Within this path, \llmbrain will return a well-formatted plan, then the execution plane will execute it to finish the task and cache it with a unique ID.  
For previous intents with IDs, the execution plane directly retrieves the cached plan and hits the fast path.
Execution on the fast path can completely bypass LLM inference, achieving native performance.   
All newly generated \agenticplans must pass a verifier to avoid possible safety issues.

The original \agenticplans are just combinations of syscall-grained tools, just for basic functionality and accuracy of classic POSIX semantics.
But it can be evolved by the \llmbrain at kernel runtime. 
The execution plane collects a predefined set of kernel metrics, which the \llmbrain uses to determine how to optimize each plan.
Meanwhile, intent-call traces are themselves kernel metrics, allowing the \llmbrain to prioritize plans for hot or critical intents.
Once an execution plan has been constructed from syscall-grained tools, the \llmbrain can asynchronously attempt to optimize the corresponding intent semantics with deeper, fine-grained tools, either reducing redundant operations or better adapting the plan to the kernel's current runtime state.

\section{Detailed Design}
\label{sec:design}

This section details designs to address challenges of performance, LLM accuracy, and safety in \sys(\S\ref{sec:intro}).
\begin{figure}[t]
  \centering
  \setlength{\belowcaptionskip}{0pt}
  \setlength{\abovecaptionskip}{2pt}
  \includegraphics[width=\linewidth]{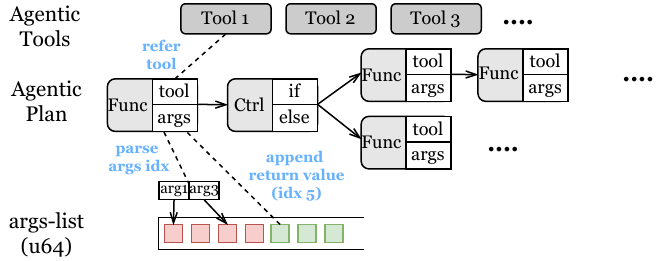}
  \caption{\textbf{The structure of an \agenticplan.} We detail the connections in the first node.}
  \label{fig:agentic-plan-structure}
\end{figure}

\subsection{\solutionperformance for Agentic Plan}
\label{sec:cache}
The performance of \sys is fundamentally limited by the latency of LLM inference. 
Current agent memory primarily optimizes token consumption, but cannot entirely bypass the LLM inference.
\sys solves it by proposing an executable and args-decoupled \agenticplan format which can be cached and reused. \llmbrain and kernel will collaborate to compile and save new plans at runtime.

\myparagraph{Executable and Args-Decoupled Agentic Plan.}
As shown in \autoref{fig:agentic-plan-structure}, an executable \agenticplan is a sequence of nodes. 
They are linked by node for execution flow control (e.g., sequence, if-else, loop). 
This format is stored with linked list structure and can be directly executed by the kernel.
What's more, \agenticplan is args-decoupled from the intent. 
Argument values are organized as an independent array.
This makes previously cached plans can be reused with different argument values. 

\isyscall's arguments can be divided into intent and args-list: 
Intent describes the purpose of the call and args descriptions, in which the args are placeholders with indexes (e.g., \codeword{arg1}) instead of actual values. 
Args-list is a value array, including actual values with unique index. 
It allocates all argument values an 8-byte slot for uniform management, which does not sacrifice functionality, as the value can be cast to pointers, handles or other primitive types. 

During plan's execution, \sys dynamically substitutes placeholders with the actual values from the args-list. Tools' return values can be appended to the args-list. Subsequent tools refer them via newly allocated index. 
If any intermediate step fails, \sys stops forward progress, reclaims allocated resources and reverts internal effects of the error tools (discussed in \S\ref{sec:micro-syscall}). After rollback, \sys returns control to user space together with the corresponding error information. 

\myparagraph{Agentic Plan JIT and Caching.}
\sys implements a JIT compiling and caching mechanism collaboratively through the \llmbrain and the kernel. 
For new intents, 
\llmbrain generates responses which follows specific schema so \llmsidecar can convert it into an \agenticplan and hand it to kernel. Each plan is cached in the kernel space with unique ID, through which user can reuse it when arguments are changed.

\subsection{\solutionfault}
\label{sec:ft-mechanism}

\sys proposes a multi-stage evolution for \semanticcompositions. 
It reduces decision space to hierarchical tools, which outperforms feeding LLMs the whole codebase or all the tools directly.
\sys adopts a \textbf{\topdowndecompose} methodology to progressively decompose coarse-grained syscall tools. 
\agentictools are organized hierarchically. Syscall tools are at the top layer, which can be further decomposed into fine-grained tools that are functionally coherent. 
Sub-tools inherit the context of their parent tools, which is reflected in descriptions for LLMs through: (1) dependency order, (2) primary functionality, (3) argument descriptions, and (4) other relations. This hierarchical representation helps the LLMs better understand the position and role of tools.

In the first stage, \llmbrain generates plans for \semanticcompositions only using syscall-grained tools, which satisfies basic functional requirements with high accuracy. 
In later stages, \sys collects the intent traces (e.g., invocation frequency, whether used by multiple processes) or kernel metrics and identifies hot or critical intents to optimize. 
Now it can evolve corresponding \agenticplans with fine-grained \agentictools for performance.
\sys uses tool search and deferred loading, progressively replacing each coarse-grained tool with a set of fine-grained tools to expose to \llmbrain, limiting the number of tools considered and thus reducing decision errors.
For severe hallucinations, \sys support a rollback solution: lets the developers review and patch cached plans offline.

\subsection{Safe Micro-Syscall and Security Guarantee}
\label{sec:micro-syscall}

One tricky challenge in \sys is how to guarantee kernel safety under malicious or incorrect plans.
Inspired by classic syscalls' security mechanism, we propose the safe micro-syscall guarantee for all fine-grained \agentictools. 
Micro-syscall is treated as self-contained API. It encapsulates complex inner logic, only exposing the necessary interface to the \llmbrain and user.
\sys export all micro-syscalls to applications. User can call them directly in arbitrary order.
Like syscalls, executing micro-syscalls does not trigger kernel panics or illegal address accesses, and failures are rolled back to ensure all-or-nothing semantics.

There are four security guarantees for micro-syscalls in \sys :

\myparagraph{Security Checking before Execution.}

We summarizes six types of syscall checking before execution: (1) user-pointer validation (UPtr), (2) existence validation (Exist), (3) length / offset / overflow validation (Len/OVF), (4) permission / capability / ownership validation (Perm), (5) resource-limit validation (RLimit), and (6) illegal-flag-combination validation (Flags).
These checks constitute the basic security guarantees for syscalls. 
Referring to them, \sys performs argument-oriented security checks before executing against injecting malicious argument values or invoking privileged operations.

\myparagraph{Indirectly Kernel Resource Access by Handles.}
Consistent with syscalls, \sys protects related kernel resources in a micro-syscall within kernel space. Users can access them only through object handles.
This design prevents attackers from directly manipulating raw kernel pointers or internal data structures, which reduces the attack surface and avoids accidental corruption. It also gives \sys a stable abstraction boundary for tracking resource lifetimes and rolling back partial effects when a semantic fails.

\myparagraph{Lock Reentrancy.}
Micro-syscall is not allowed to retain global kernel locks across boundaries. Any lock acquired during its execution must be released before returning, regardless of successes or failures. This rule prevents subsequent tool invocations from inheriting hidden lock state, avoiding cross-node deadlocks and making each micro-syscall reentrant as an independent execution unit.


\subsection{Static Verifier}
\label{subsec:static-verifier}
\sys provides a static verifier to check bounded loops.
\sys guarantees plans terminate within a finite number of steps, preventing \isyscall from hanging indefinitely in the kernel for a given \semanticcomposition. 
To enforce this property, \agenticplan adopts a constrained loop structure in which each loop-control group is separated from tool arguments and statically specified~\cite{298732}.  
When new \agenticplan is created, the verifier checks loop-control groups to ensure that its initial condition and iteration rule can reach termination condition.




\section{\sys Implementation}
\label{sec:case-study}

Without loss of \arch's generality across different OSes, we implemented three sets of \agentictools on Linux in process management, file systems, and memory management domain respectively.

This section focuses on the overview of implementations in \sys.
It then provides the overview of a set of fine-grained \agentictools in process management, and shows optimized \agenticplan evolved by LLMs.

\subsection{Implementation Overview}

\myparagraph{Invocation Interfaces.}
\sys gives three isyscall-related interfaces. Applications can use \codeword{isyscall\_rt} to submit intents, allowing \llmbrain to construct \agenticplans at runtime. They can also use \codeword{isyscall\_dev} to submit well-designed plans, which are executed directly by kernel. In addition, \codeword{isyscall\_dry\_run} is proposed to construct new plan from an intent without actual invocation, which helps to review and refine intents.

\myparagraph{Session Split in Forking.} To solve the issue of split LLM contexts of child processes, 
\sys prepares separate sessions for them with mechanism below:   
(1) \textbf{Child Reentrancy:} We set callback function in \codeword{ret\_from\_fork}. When new child process is awakened, the callback function will be invoked to continue interacting with \llmbrain.
(2) \textbf{Multi-Session Cloning:} \llmsidecar will create a new session for the child process. The new session has its own state that is independent from the parent. The child enters the session with original intent prompt, but specifies that \llmbrain only focuses on the child-related intent.
(3) \textbf{Collecting Child Agentic Plans:} A complete \agenticplan will have both parent plan and child plan. 
\llmsidecar allocate child\_id for each child process. The parent session will poll and collect all child plans to construct complete \agenticplan. 

\myparagraph{Accessibility of User Space State.}
To maintain accessibility of data and state (e.g., intent, args) across processes,  
\sys passes them to kernel space when plan is executed, reverting the responsibility of state preservation from user space to kernel space. 
This operation doesn't incur additional overhead, as there are necessary copy operations if kernel wants to access user space data by pointers. 
For inter-process data, \sys creates an initial copy when creating a new process, then fully hands over its lifecycle to the child. 

\subsection{Optimized Syscall Composition: Fork + Exec}
\begin{table}[t]
\centering
\scriptsize
\caption{\textbf{Fork and Exec Fine-grained Tools on \sys.}}
\label{tab:linux-fork-exec-tools}
\begin{tabular}{p{1.2cm}lp{3cm}}
\toprule
\textbf{Category} & \textbf{Agentic Tool Interface} & \textbf{Function Descriptions} \\
\midrule
\multirow{6}{1.2cm}{\textbf{Process Create}} & \texttt{init\_ctx} & Init operation context \\
& \texttt{copy\_process} & Create child process \\
& \texttt{capture\_pid} & Capture child PID \\
& \texttt{set\_child\_sched\_func} & Bind reentrant entry for child\\
& \texttt{wake\_child\_and\_report} & Wake up and selectively wait child, then report trace event \\
& \texttt{ctx\_finalize} & Clean and finish context \\
\midrule
\multirow{3}{1.2cm}{\textbf{Execve}} & \texttt{run\_exec} & Exec with user args \\
& \texttt{run\_exec\_consuming\_kargs} & Consume kernel args to Exec \\
& \texttt{execve\_free\_kargs} & Free kernel args \\
\midrule
\multirow{2}{1.2cm}{\textbf{Trace}} & \texttt{pick\_trace} & Select ptrace event type \\
& \texttt{trace\_sched\_fork} & Set fork trace event \\
\midrule
\textbf{Others} & \texttt{copy\_pid\_to\_user} & Write child pid (or 0) to user \\
\bottomrule
\end{tabular}
\end{table}

\myparagraph{Introduction of Fork/Exec Micro-Syscalls.} 
As shown in Table~\ref{tab:linux-fork-exec-tools},
\codeword{fork/vfork} functions are decomposed into two Layer 2 function categories, namely ``Process Create'', and ``Trace''. \codeword{Execve} function does not require further decomposition, only adding a set of wrapper tools that directly use \codeword{kargs}. One helper tool is needed to copy the pid value back to user space to preserve fork semantics.
Each tool has its own arguments and return values. 
For example, \codeword{copy\_process} can decide whether to share \codeword{mm\_struct} and \codeword{sighand} by corresponding arguments. 

\myparagraph{Exploring Appropriate Agentic Plan.}
\begin{figure}
  \centering
  \includegraphics[width=1.0\linewidth]{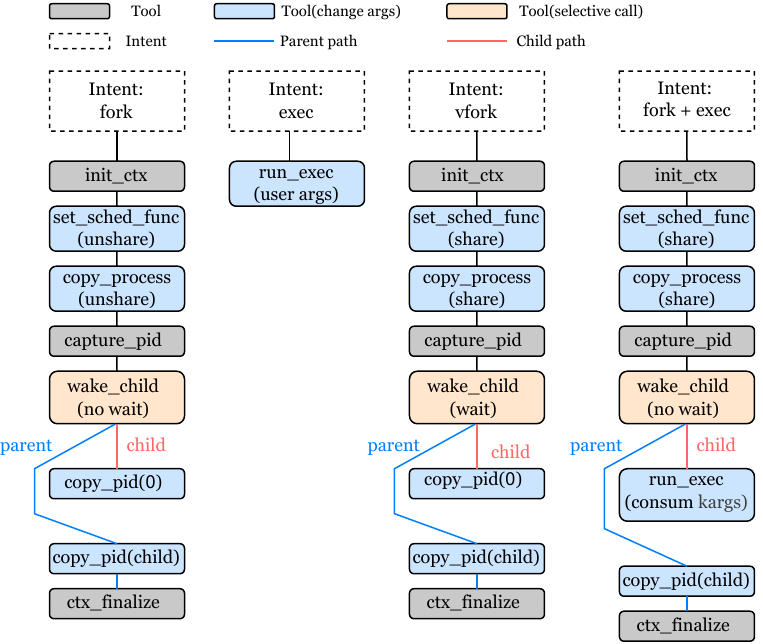}
  \setlength{\belowcaptionskip}{-15pt}
  \setlength{\abovecaptionskip}{-5pt}
  \caption{\textbf{Overview of the composition of \textit{fork and exec} tools.} Child processes can re-enter sessions and communicate with \llmbrain. For different intents, some tools have different arguments, and partial operations in some tools are selectively called.}
  \label{fig:agentic-plans-fork-exec}
\end{figure}

We prepare four intents to explore appropriate \agenticplans under different scenarios: (1) fork (2) exec (3) vfork (4) fork and exec immediately. The first three intents are classic syscall semantics in Unix-like systems. 
The last intent is a \semanticcomposition in one kernel trap.
\autoref{fig:agentic-plans-fork-exec} shows the overview of LLM's \agenticplans.
Fine-grained fork/vfork tools always begin with the \codeword{init\_ctx} micro-syscall, 
Whether the execution terminates normally or fails midway, \codeword{ctx\_finalize} is invoked to clear the context state. 

Intent~(4) demonstrates that \sys and its fine-grained micro-syscalls enable the LLM to generate optimized plan for \semanticcomposition : referring to vfork paradigm, \sys creates child process with shared resources (e.g., \codeword{mm\_struct}, \codeword{sighand\_table}) to reduce memory copying. 
What's more, the child finishes all operations in kernel space,
so parent can skip vfork waiting without access contention on user memory. 
This validates the expressiveness of our decomposition methodology: the micro-syscalls are fine-grained enough for optimization, yet semantically clear enough for automated composition.

\section{Evaluation}

The evaluation of \sys includes reasoning accuracy, compatibility, and performance optimization of \semanticcompositions.
We evaluate \sys's reasoning accuracy in Linux-x86 on three domains: memory management, file system, and communication, each equipped with ten representative intents.
A compatibility test is conducted for the same intents under four Linux kernel versions.
The performance of optimized \agenticplans is evaluated on three cases: \forkexec, \batchunlink, and \slidingwindowmmap. 
The structure of the evaluation is as follows:
\begin{itemize}[leftmargin=2.5mm,topsep=2pt]
    \item Testbed and performance test cases. (\S\ref{subsec:eval-testbed}, \S\ref{subsec:eval-testcases})
    \item Accuracy test to show the improvement of \sys's reasoning accuracy under multiple intents compared with direct coding by \llmbrain. (\S\ref{subsec:eval-accuracy})
    \item Micro-benchmark to show \agenticplan's ideal optimization result, and integration of \sys's \isyscall into applications for e2e performance evaluation. (\S\ref{subsec:eval-micro}, \S\ref{subsec:eval-app})
    \item Intent compatibility test under four Linux versions. (\S\ref{subsec:eval-compatibility})
    \item Latency breakdown test of \isyscall and ablation study to prove the necessity of \solutionperformance. (\S\ref{subsec:eval-JIT-necessity})
\end{itemize}

\subsection{Environment Setup for Evaluation}
\label{subsec:eval-testbed}
The testbed of evaluation is as follows.
CPU: Intel(R) Xeon(R) Platinum 8468V with 2.4 GHz, 192 Cores. Memory: 1024 GB DDR5 ECC @ 4800 MHz (32 × 32 GB, 2R, 1.1 V). Host OS: Ubuntu 22.04 with Linux-6.18.6. Virtualization Platform: QEMU-10.2.0-KVM. Except for compatibility test, all performance tests are conducted in bare-metal environment.
We select four models as the \llmbrain: DeepSeek-V4-Pro, Qwen3.6-Plus, Gemini-3.1-Pro-Preview, and GPT-5.4. 

\subsection{Use Cases of Performance Tests}
\label{subsec:eval-testcases}
Beyond the \textit{fork + exec} in \S\ref{sec:case-study}, we further provide two more sets of tools suited for different scenarios to show the generality and scalability of \sys's \agenticplans.

\myparagraph{Fork+Exec.} 
New semantic enables the shared-resource options during \codeword{copy\_process} without following synchronous waiting, which eliminates overheads of redundant memory copying and blocking waiting.
We integrate optimized \forkexec semantic into the Busybox-ash shell.
When shells need to launch child processes running in the background, 
optimized semantic can effectively reduce the latency compared with traditional \codeword{fork} and \codeword{execve}, thus improving the user interaction experience.     

\myparagraph{Batch Unlink.}  
During the deletion of multiple files under the same directory.
New semantic acquires the parent directory's inode lock just once and batches the metadata updates, significantly reducing the overhead associated with repeated lock contention and redundant filesystem journaling.
We integrate optimized \batchunlink semantic into the Unix-rm. 
In the compile flow, rm always needs to clean many intermediate build artifacts under the same parent directory (e.g., Redis, Coreutils). These repos always build their core binary files directly under the \textit{/src} directory, and thus are pretty suitable for \batchunlink.

\myparagraph{Sliding Window Mmap.} 
\sys addresses the page cache pollution problem in background analytics by introducing the \slidingwindowmmap semantic. 
When processing large files in a memory-pressure environment, standard \codeword{mmap} or \codeword{read} operations can easily evict high-priority page cache belonging to foreground service, causing performance degradation.
The \slidingwindowmmap semantic maintains a small, fixed-size sliding window (typically 16 MB) as page cache for the large file. When the process accesses data outside the current window, the tool triggers a page fault that zaps the old mappings and slides the window to a new offset. 
We integrate it into the Unix-grep.
Evaluation demonstrates that grep with \slidingwindowmmap eliminates the memory content performance impact on the front-end file server (e.g., Nginx).

\begin{figure}[tb]
  \centering
  \includegraphics[width=0.99\linewidth]{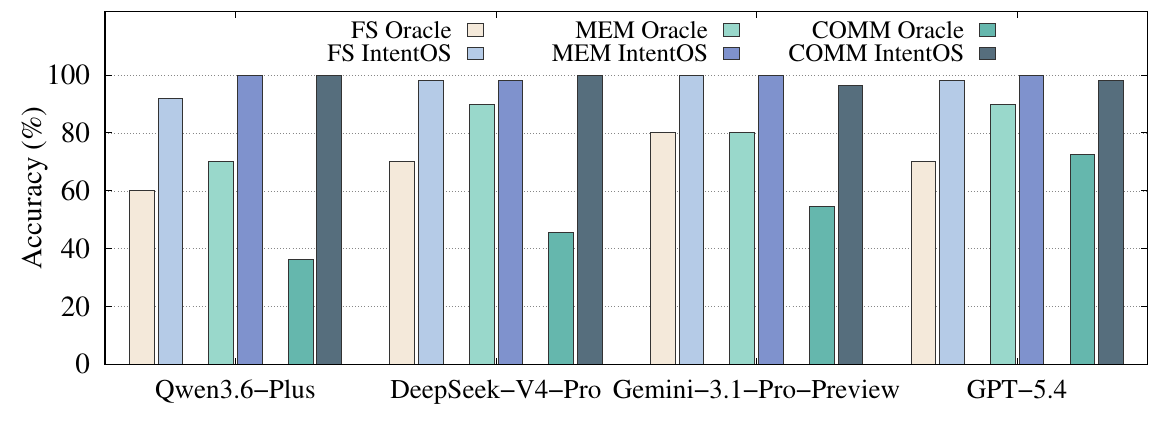}
  \setlength{\belowcaptionskip}{-10pt}
  \setlength{\abovecaptionskip}{-10pt}
  \caption{\textbf{\llmbrain Accuracy test.} It includes four models in three domains. Oracle: receives kernel source to generate new syscall; \sys: agentic kernel tools. FS/MEM/COMM: file system, memory, and communication.}
  \label{fig:accuracy-test}
\end{figure}

\subsection{Accuracy Test for \sys}
\label{subsec:eval-accuracy}
We evaluate the accuracy of \sys compared with naive agent coding paradigm.
The evaluation covers four models across three domains.
Each domain contains about ten representative intents for various \semanticcompositions, for a total of 31 intents.
We use the same intent prompts for two paradigms.
In the oracle baseline, we directly give the \llmbrain the kernel code and ask it to generate new syscalls.
In \sys, \llmbrain gets the syscall-grained micro-syscalls and generates \agenticplans.
As shown in \autoref{fig:accuracy-test}, averaged over all models and intents, \sys achieves an average accuracy of 98.38\%, while the oracle baseline achieves 66.74\%.
It demonstrates that the effectiveness to narrow the reasoning space into tool sequences.
In contrast, the oracle baseline shows that raw kernel code is too complex for accurate reasoning, validating our claim that decomposition into self-contained micro-syscalls is essential for runtime composability.
More importantly, unlike the oracle baseline, \sys does not require recompilation to enable new compositional semantics.

\begin{figure}[htb]
  \centering
  \setlength{\abovecaptionskip}{6pt}
  \setlength{\belowcaptionskip}{-12pt}
  \begin{minipage}[t]{0.48\linewidth}
      \centering
      \includegraphics[width=\textwidth]{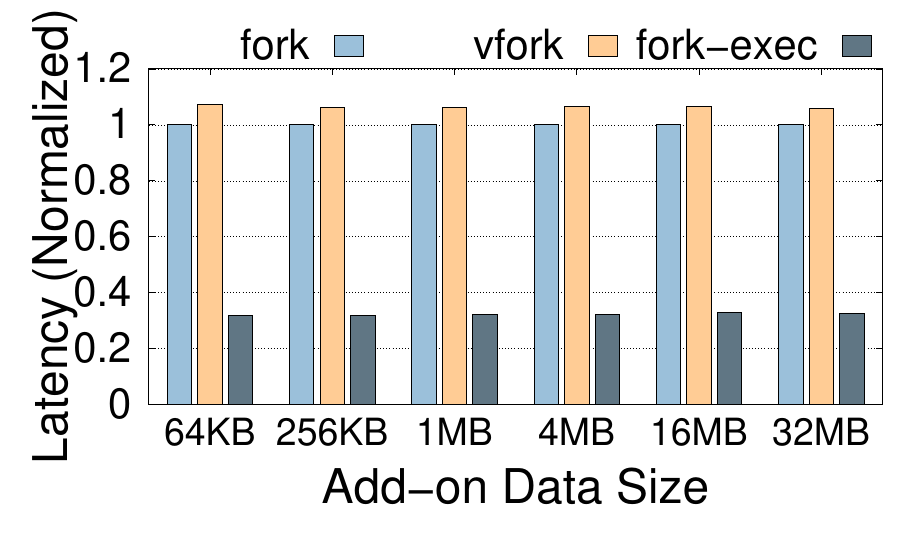}
      \footnotesize
      \textbf{(a) Micro-benchmark.}
  \end{minipage}
  \hfill
  \begin{minipage}[t]{0.48\linewidth}
      \centering
      \includegraphics[width=\textwidth]{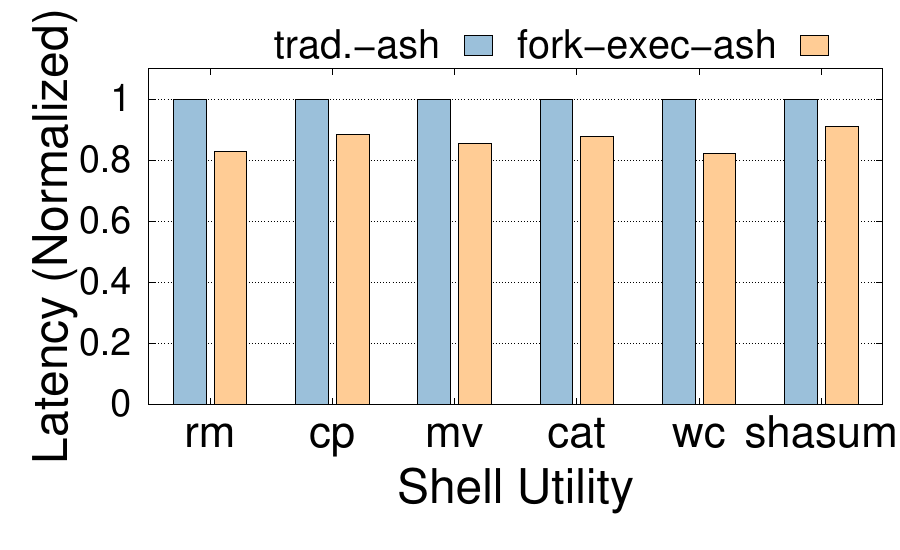}
      \footnotesize
      \textbf{(b) ash Shell.}
  \end{minipage}
  \caption{\textbf{Performance of Agentic Fork \& Exec.} (a) shows micro-benchmark latency; (b) shows end-to-end latency for ash shell. The original file size in (a) is 880KB. We attach extra data segment to increase the size. Data in each group is normalized.}
  \label{fig:eval-fork-exec}
\end{figure}

\begin{figure}[htb]
  \centering
  \setlength{\abovecaptionskip}{6pt}
  \setlength{\belowcaptionskip}{-12pt}
  \begin{minipage}[t]{0.48\linewidth}
      \centering
      \includegraphics[width=\textwidth]{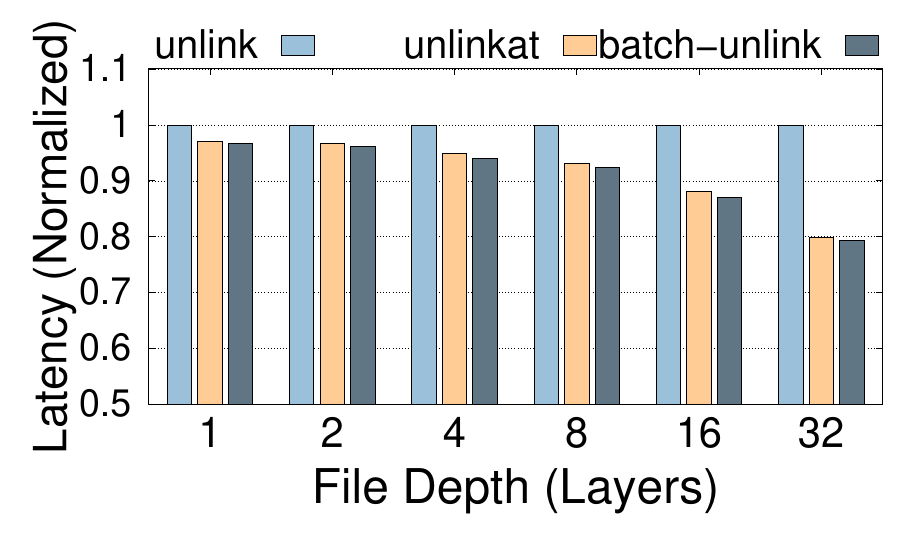}
      \footnotesize
      \textbf{(a) Micro-benchmark.}
  \end{minipage}
  \hfill
  \begin{minipage}[t]{0.48\linewidth}
      \centering
      \includegraphics[width=\textwidth]{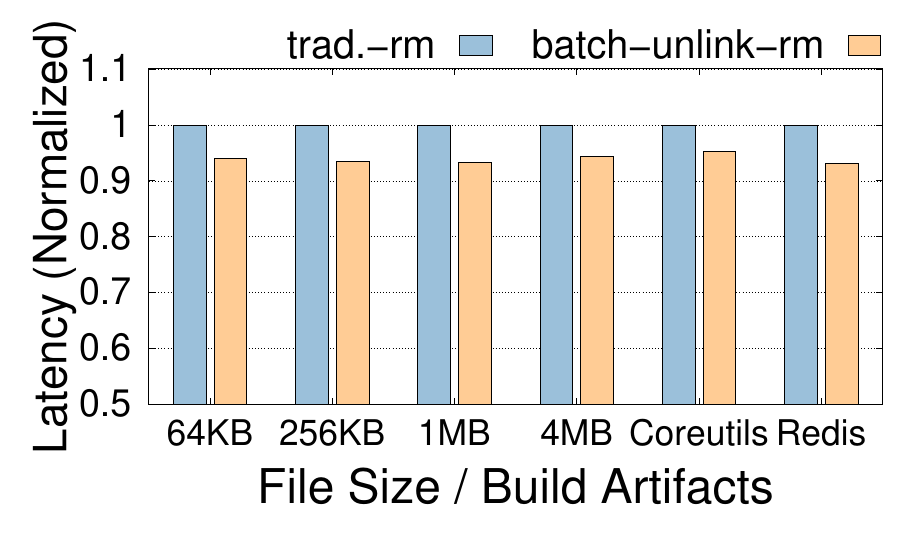}
      \footnotesize
      \textbf{(b) rm Utility.}
  \end{minipage}
  \caption{\textbf{Performance of Agentic Batch Unlink.} (a) shows micro-benchmark latency with different file depth; (b) shows end-to-end latency for rm utility. The number of deleted files is 50 except for Coreutils and Redis which clean real-world build artifacts. Data in each group is normalized.}
  \label{fig:eval-batch-unlink}
\end{figure}

\begin{figure}[htb]
  \centering
  \setlength{\abovecaptionskip}{6pt}
  \setlength{\belowcaptionskip}{-12pt}
  \begin{minipage}[t]{0.485\linewidth}
      \centering
      \includegraphics[width=\textwidth]{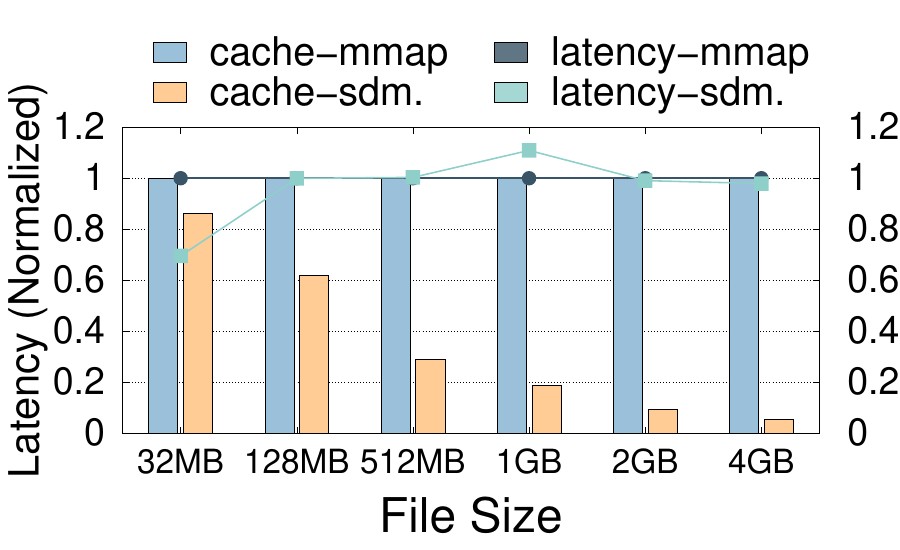}
      \footnotesize
      \textbf{(a) Micro-benchmark.}
  \end{minipage}
  \hfill
  \begin{minipage}[t]{0.485\linewidth}
      \centering
      \includegraphics[width=\textwidth]{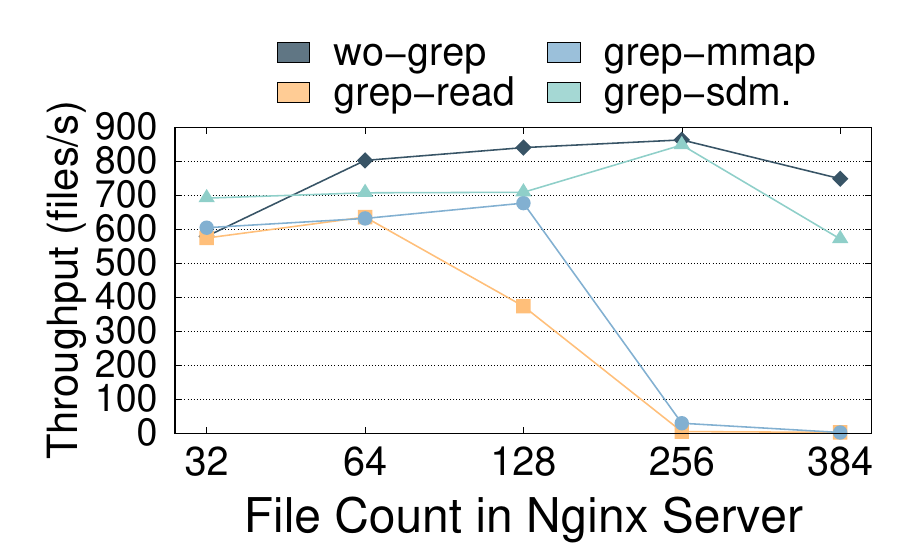}
      \footnotesize
      \textbf{(b) Nginx and grep.}
  \end{minipage}
  \caption{\textbf{Performance of Agentic Mmap Overlap.} (a) shows micro-benchmark latency and page cache size; (b) shows Nginx throughput with memory contention. \textit{sdm.} refers to sliding mmap. In memory-pressure scenario, available memory is restricted to 5GB. Each Nginx-served file is 4MB. Data in each group is normalized.}
  \label{fig:eval-mmap-overlap}
\end{figure}

\subsection{Performance Micro-benchmark}
\label{subsec:eval-micro}
We first evaluate the ideal performance of each semantic powered by \sys's \agenticplans through micro-benchmarks. 

\myparagraph{Fork + Exec.} As shown in \autoref{fig:eval-fork-exec}(a), the \agenticplan outperforms the traditional sequences under different executable file sizes. It is worth noting that while \codeword{vfork} saves memory copying overhead compared to \codeword{fork}, its synchronous waiting mechanism sometimes causes increased latency. \sys overcomes this and achieves a latency reduction of up to 69.83\% compared to the separate \codeword{fork} and \codeword{exec} baseline.

\myparagraph{Batch Unlink.} The micro-benchmark for \batchunlink (\autoref{fig:eval-batch-unlink}(a)) demonstrates the efficiency of batch unlinking files in same directory with different directory depths. By reducing path resolution overhead, the \agenticplan reduces file deletion overhead by at most 20.69\% over the original iterative \codeword{unlink}. It also outperforms \codeword{unlinkat} by acquiring directory locks and committing metadata changes fewer times.

\myparagraph{Sliding Window Mmap.} \autoref{fig:eval-mmap-overlap}(a) compares the performance and resource consumption of the sliding window mechanism against standard \codeword{mmap}. While keeping comparable data access latency with original \codeword{mmap} (at most 10.89\% additional overhead), it significantly reduces the page cache memory footprint by up to 94.49\% through its fixed-size window and Direct I/O path, effectively preventing tasks from exhausting system memory resources.

\subsection{Integration into Real-World Applications}
\label{subsec:eval-app}
We further demonstrate the practical benefits of \sys by integrating \agenticplans into applications.

\myparagraph{ash Shell (for Fork + Exec).} We integrate the \forkexec into the Busybox-ash shell for launching various background jobs. As illustrated in \autoref{fig:eval-fork-exec}(b), the end-to-end latency for job submission is reduced by up to 17.55\%. This improvement is vital in shell scripts that launch many backend-lived processes, as the reduction in return latency can directly translate to better shell responsiveness.

\myparagraph{rm Utility (for Batch Unlink).} We evaluate the performance of deleting multiple sub-files with different sizes, and use agentic rm to clean the Redis / Coreutils build artifacts under the \textit{src/} directory. \autoref{fig:eval-batch-unlink}(b) shows that the agentic \batchunlink reduces the total execution time by 6.97\%, showcasing the advantage of reducing lock contention and metadata update overhead in filesystem operations.

\myparagraph{Nginx + grep (for  Sliding Mmap).} The benefit of \slidingwindowmmap is observed in memory-pressure scenario. As shown in \autoref{fig:eval-mmap-overlap}(b), compared to other grep methods, it avoids severe page cache contention pressure and performance degradation of frontend Nginx. As the number of files served by Nginx increases, \slidingwindowmmap causes at most 75.52\% lower performance impact than other grep methods.

\begin{figure}[tb]
  \centering
  \begin{minipage}[t]{0.48\linewidth}
      \centering
      \includegraphics[width=\textwidth]{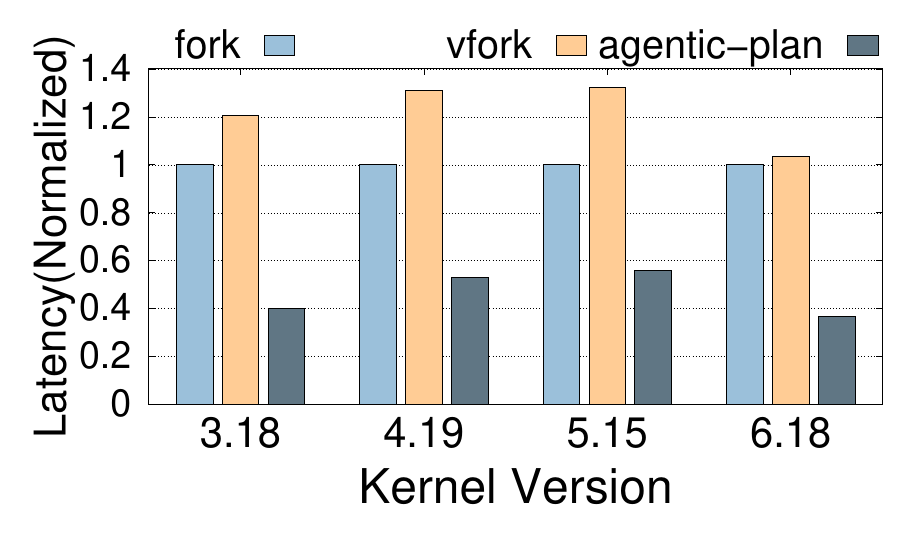}
      \footnotesize
      \textbf{(a) Fork \& Exec.}
  \end{minipage}
  \hfill
  \begin{minipage}[t]{0.48\linewidth}
      \centering
      \includegraphics[width=\textwidth]{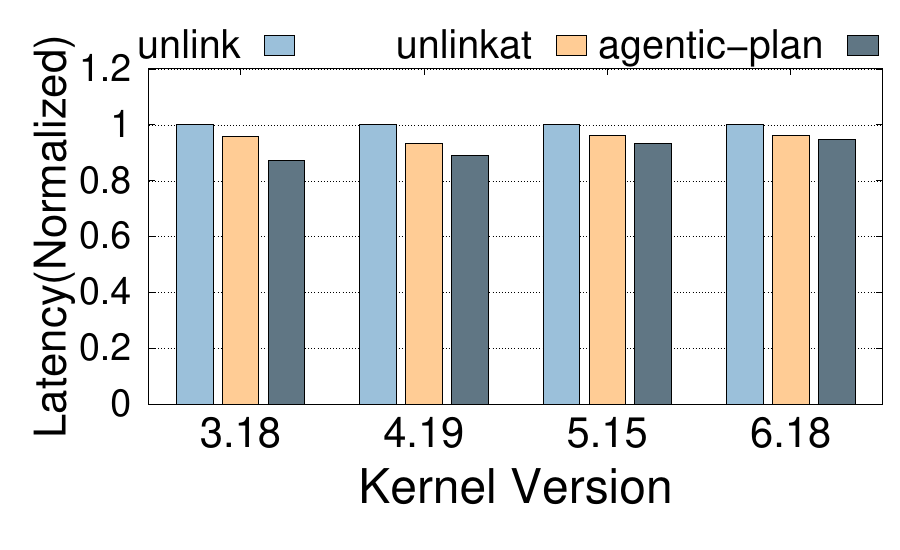}
      \footnotesize
      \textbf{(b) Batch Unlink.}
  \end{minipage}
  \caption{\textbf{Compatibility tests across Linux versions.} We evaluate micro-syscall-based \agenticplans on four kernel versions. The intent prompts are the same in each domain.}
  \label{fig:compatibility-test}
\end{figure}

\subsection{Intent Compatibility Evaluation}
\label{subsec:eval-compatibility}
We evaluate the compatibility of \sys across four popular Linux kernel versions: 3.18, 4.19, 5.15, and 6.18, in a QEMU-KVM environment. For fair comparison, we use the same intents as those in the micro-benchmark evaluation in \S\ref{subsec:eval-micro}. 
As shown in \autoref{fig:compatibility-test}, even though underlying tool implementations differ across kernel versions, \sys can still compose the required optimized plans and consistently deliver performance improvements. 
This result demonstrates \sys's compatibility by decoupling intent from mechanism, allowing semantic construction to remain stable despite kernel API changes, accelerating the evolution of system.

\subsection{Necessity of \solutionperformance}
\label{subsec:eval-JIT-necessity}
\begin{table}[t]
\caption{\textbf{Ablation study on \agenticplan.} The end-to-end latency of \isyscall with and without cached \agenticplans.}
\centering
\footnotesize
\renewcommand{\tabularxcolumn}[1]{m{#1}}
\begin{tabularx}{\linewidth}{l*{3}{>{\centering\arraybackslash}X}}
\toprule
 & \textbf{fork+exec} & \textbf{batch unlink} & \textbf{sliding mmap} \\
\midrule
\textbf{w/o plan}   & 56340 ms & 26477 ms  & 28816 ms \\
\textbf{with plan}  & 4.56 ms  & 4.40 ms   & 8742 ms  \\
\midrule
\textbf{Speedup}    & 12378$\times$ & 6020$\times$ & 3.30$\times$ \\
\bottomrule
\end{tabularx}
\label{tab:ablation-plan}
\end{table}

The necessity of \solutionperformance is evaluated on three test cases, which aims to prove that: (1) bypassing inference latency is critical for performance. (2) The overhead of the \agenticplan execution framework is negligible.

As shown in \autoref{tab:ablation-plan}, the use of cached \agenticplans significantly reduces the end-to-end latency of \isyscall. 
Without cached plans, the system must perform full reasoning for each intent on critical path, which introduces unacceptable overhead. 
By bypassing the inference step for previous intents with JIT method, \sys achieves a speedup of up to 12378$\times$.
Overhead of \slidingwindowmmap is high because of the page fault configuration operations on large memory space.
But it still provides a 3.3$\times$ speedup when using cached plans.  

\begin{figure}[t]
  \centering
  \includegraphics[width=\linewidth]{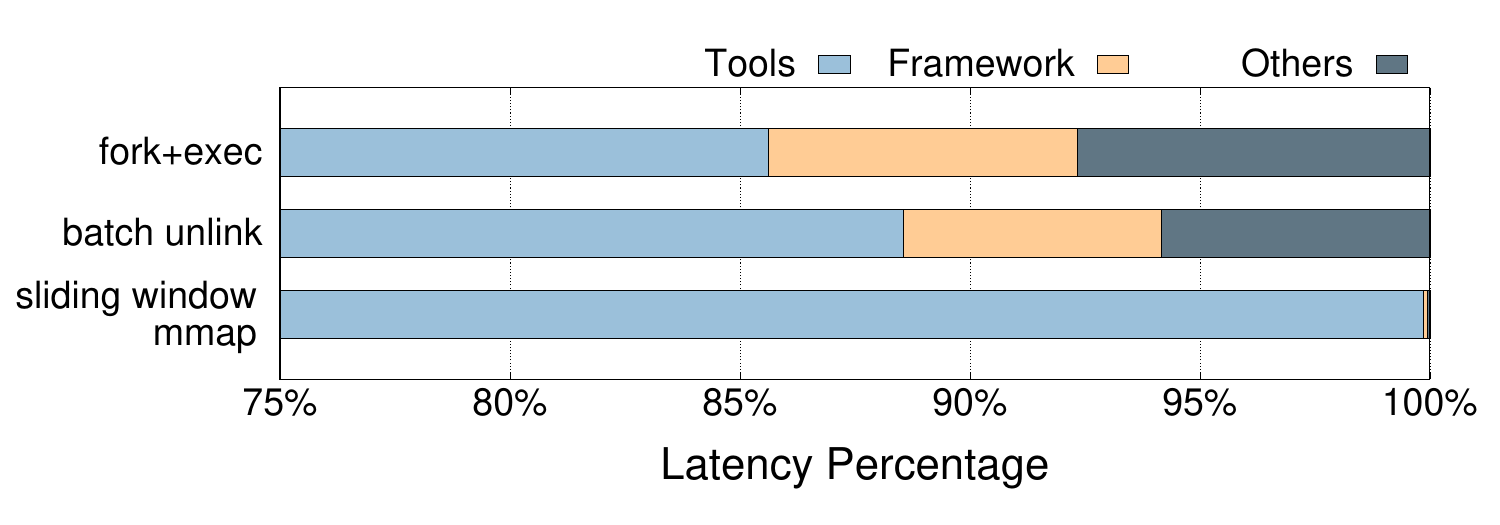}
  \caption{\textbf{Latency breakdown of isyscall in different \agenticplans,} including the execution of tools (Tools), additional overhead of \agenticplan's framework (Framework), and other overheads (Others).}
  \label{fig:eval-breakdown}
\end{figure}

We also conduct a breakdown analysis of \isyscall.
We decompose the whole latency into three components: (1) The regular execution of functional tools. (2) The additional overhead of \agenticplan's execution framework, which includes plan retrieval, argument parsing from indexes, extra pointer accesses and so on. (3) Other overheads such as context switches. 
The test result is shown in \autoref{fig:eval-breakdown}. The percentage of framework latency in \agenticplan execution is up to 6.714\%.
It demonstrates that the additional overhead of \agenticplan's framework is negligible.

\section{Discussion}

\myparagraph{When \sys Is Worthwhile.}
\sys is not intended to replace mechanisms such as \texttt{posix\_spawn}, \texttt{vfork}, \texttt{io\_uring}, \texttt{madvise}, or eBPF. 
Instead, it complements fixed OS interfaces in workloads with recurring intents and large semantic composition spaces. 
Developers may also directly construct semantics over micro-syscalls, while \llmbrain provides a faster and more flexible synthesis path.

\myparagraph{The Role of \llmbrain in \sys.}
\sys seeks only to leverage LLM capabilities for orchestration; the LLM runtime environment need not be ported into the kernel.
\sys explores how LLM capabilities can support kernel runtime specialization, but the current design remains limited by cold-path latency and probabilistic errors on unseen intents. 
We therefore view \llmbrain as a mechanism for improving flexibility, reducing developer burden, and expanding runtime adaptation currently.

\myparagraph{Expand \sys to Other OS Architecture.}
Although prototyped on Linux, the agentic kernel is highly feasible across diverse OS architectures. In microkernels (e.g., seL4~\cite{10.1145/2517349.2522720,klein2009sel4}), \agentictools seamlessly map to native kernel object methods or user-space IPC endpoints. In modular BSDs, structured interfaces like VFS vnode operations can be directly wrapped as composable tools. Building on this, \arch can deconstruct these tools into granular, atomic ones such as isolating state queries, permission validations, and execution triggers.

\section{Related Work}
\label{sec:related}

\myparagraph{Optimizations for syscalls or the compositions.}
Userspace Bypass (UB)~\cite{288604} accelerates syscall-intensive apps by moving userspace instructions and syscall invocations into kernel.
Privbox~\cite{280708} accelerates syscalls by allowing syscall-intensive code to run in a semi-privileged yet sandboxed execution mode, thereby reducing invocation overhead while preserving the familiar synchronous, function-like syscall interface. 
BASTION~\cite{10.1145/3582016.3582066} protects syscall integrity by enforcing constraints at runtime through compiler and monitor support. 
However they do not introduce semantic optimizations for \semanticcomposition.

\myparagraph{Customized Kernel and Kernel Bypass Method.}
Customized-kernel and kernel-bypass approaches address the rigidity of traditional kernels in two ways. Exokernels and modern bypass frameworks such as DPDK~\cite{engler1995exokernel, dpdk} expose lower-level resources to applications and often rely on LibOS-style software stacks to build specialized abstractions; DemiKernel similarly keeps only limited control logic in the kernel while offloading data-path operations to a LibOS tailored to specific hardware~\cite{zhang2019m}. 
In contrast, customized-kernel mechanisms like eBPF allow developers to safely inject programmable logic into the kernel~\cite{308758, 280870, 10.1145/3731569.3764820, 298730, 10.1145/3731569.3764796, 10.1145/3694715.3695950, 10.1145/3689031.3717497,10.1145/3689031.3696094, 10.1145/3627703.3629562, 308448, 308452, 298490, 298531, 10.1145/3620666.3651387, cilium}. 
These systems improve flexibility for specific mechanisms, but shift complexity to manual development and do not enable runtime composition of new syscall-level abstractions.
Related works also support customized scheduling logic~\cite{10.1145/3477132.3483548, 10.1145/3477132.3483542}. However, they focus on scheduling policies rather than constructing new kernel semantics.

\myparagraph{LLM Generation or Machine Learning for Systems.}
Recent work uses LLMs for system code generation and evolution. SYSSPEC generates and evolves file systems from high-level specifications~\cite{liu2026sharpenspeccutcode}, while ECO refactors large-scale production code using optimization patterns~\cite{lin2025ecollmdrivenefficientcode}.
And SysGPT~\cite{308736} delivers context-aware, research-aligned optimization suggestions.
These efforts mainly target static coding during the development cycle. In contrast, \sys focuses on runtime system semantic construction.
Many works focus on machine learning for systems~\cite{10.1145/3767295.3803589, 10.1145/3689031.3717496, liao2024fairefficientlearningbasedcongestion, 308612, 10.1145/3575693.3575697, 10.1145/3503222.3507704, 10.1145/3445814.3446693, 10.1145/3575693.3575714}. \sys explores LLM agentic paradigm in kernel, and is orthogonal to these efforts.

\myparagraph{Other Related OSes.}
Declarative approaches, because of the benefits, are already used in prior OS systems.
For example, Barrelfish~\cite{10.1145/2110356.2110361} and Singularity~\cite{10.1145/1217935.1217941} utilize a declarative approach to configure (PCI) devices and implement drivers.
HeliOS~\cite{nightingale2009helios} uses a declarative affinity to describe application placement requirements.
DBOS~\cite{10.14778/3485450.3485454} reuses DBMS and SQL (a declarative language) to manage OS states.
Even Linux allows applications to advise on memory usage using the \codeword{madvise}.
Compared to these systems, \sys is the first OS architecture to treat \emph{declarative} kernel semantics extension as a core design principle for optimization and application adaptability.

\section{Conclusion}
\label{sec:concl}

This paper proposes \arch, a new kernel architecture, and its Linux-based prototype, \sys, which decomposes syscall functions into fine-grained, composable kernel-native micro-syscalls and enables their runtime re-orchestration to construct optimized system semantics.
Extensive evaluations show that \sys not only maintains high compatibility and accuracy but also delivers significant performance gains for real-world applications.
We believe this work demonstrates a promising path toward runtime-composable kernel semantics, where micro-syscall decomposition provides the architectural foundation and LLMs serve as the practical enabler for automated plan generation.

\bibliographystyle{ACM-Reference-Format}
\bibliography{references}

\clearpage
\appendix

\end{document}